# The MIRI Medium Resolution Spectrometer Calibration Pipeline

A. Labiano*[a], R. Azzollini[b], J. Bailey[c], S. Beard[d], D. Dicken[e], M. García-Marín[fg], V. Geers[d], A. Glasse[d], A. Glauser[a], K. Gordon[g], K. Justtanont[h], P. Klaassen[d], F. Lahuis[i], D. Law[g], J. Morrison[gj], M. Müller[k], G. Rieke[j], B. Vandenbussche[l], G. Wright[d]

[a]Institute for Astronomy, ETH Zurich, Wolfgang-Paulistrasse 27, CH-8093 Zurich, Switzerland. [b]Dublin Institute for Advanced Studies, 31 Fitzwilliam Place, Dublin 2, Ireland. [c]Leiden Observatory, Leiden University, 2333CA Leiden, The Netherlands. [d]The UK Astronomy Technology Centre, Blackford Hill, Edinburgh, EH9 3HJ, UK. [e]Service d'Astrophysique, CEA-Saclay, Gif-sur-Yvette 91191, France. [f]European Space Agency. [g]Space Telescope Science Institute, 3700 San Martin Drive, Baltimore, MD 21218, USA. [h]Chalmers University of Technology, Onsala Space Observatory, S-439 92 Onsala, Sweden. [i]SRON Netherlands Institute for Space Research, P.O. Box 800, 9700 AV Groningen, The Netherlands. [j]Steward Observatory, University of Arizona, Tucson, AZ, 85721. [k]Kapteyn Astronomical Institute, Groningen, 9700 AV, The Netherlands. [l]KU Leuven, Institute of Astronomy, Celestijnenlaan 200d - Box 2401, 3001 Leuven

*alvaro.labiano@phys.ethz.ch

## ABSTRACT

The Mid-Infrared Instrument (MIRI) Medium Resolution Spectrometer (MRS) is the only mid-IR Integral Field Spectrometer on board James Webb Space Telescope. The complexity of the MRS requires a very specialized pipeline, with some specific steps not present in other pipelines of JWST instruments, such as fringe corrections and wavelength offsets, with different algorithms for point source or extended source data. The MRS pipeline has also two different variants: the *baseline* pipeline, optimized for most foreseen science cases, and the *optimal* pipeline, where extra steps will be needed for specific science cases. This paper provides a comprehensive description of the MRS Calibration Pipeline from uncalibrated slope images to final scientific products, with brief descriptions of its algorithms, input and output data, and the accessory data and calibration data products necessary to run the pipeline.

**Keywords:** JWST, MIRI, MRS, Calibration, Pipeline, Data.

## 1. INTRODUCTION TO THE MRS

The Medium Resolution Spectrometer (MRS) is the Integral Field Spectrometer of the JWST Mid-Infrared Instrument (MIRI)[1,2,3]. The MRS is designed to cover the wavelength range from 5 to 28.5 μm at spectral resolving powers of R~3000. It consists of four integral field units (IFU), covering four different spectral channels in one single observation. Channel 1 corresponds to the shortest wavelengths, and channel 4 to the longest wavelengths. Two grating and dichroic wheels (GDW) select the wavelength coverage within all four channels simultaneously, dividing each channel into three spectral sub-bands: "SHORT" (or "A"), "MEDIUM" (or "B") and "LONG" (or "C"). Thus, to obtain a spectrum for the whole 5 to 28.5 μm range, the observer needs to stitch together 12 sub-band spectra, obtained in three exposures (each exposure with a different GDW position): 1A-2A-3A-4A (exposure 1), 1B-2B-3B-4B (exposure 2), and 1C-2C-3C-4C (exposure 3). Figure 1 illustrates the wavelength coverage of the MRS. A full description of the MRS is given in [4].

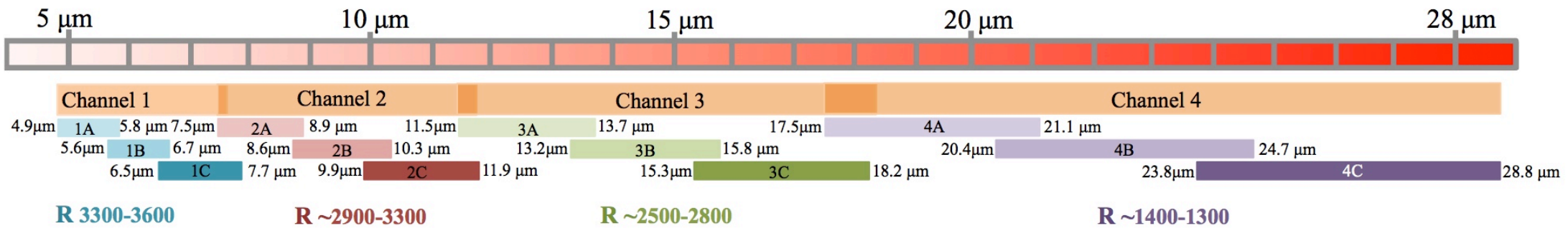

Figure 1 – Wavelength coverage of the different channels and sub-bands of the MRS

# 2. THE MRS PIPELINE

A high-level diagram of the MIRI calibration pipeline is presented in Figure 2, with the MRS specific steps highlighted in red. The MRS Calibration Pipeline processes LEVEL 2a data (i.e., uncalibrated slopes images)[4], to produce LEVEL 2b data (calibrated slopes images), and MRS LEVEL 3 data (spectral cubes and 1D spectra) for each observation or association (a set of exposures that are processed together to produce a single LEVEL 3 data product such as mosaics, different visits to the same target, etc.). This process takes place in two sets of steps. The first step is called CALSPEC2, and brings data from LEVEL 2a to LEVEL 2b. The second step, CALSPEC3 (formerly named CALIFU3), brings the data from LEVEL 2b to LEVEL 3.

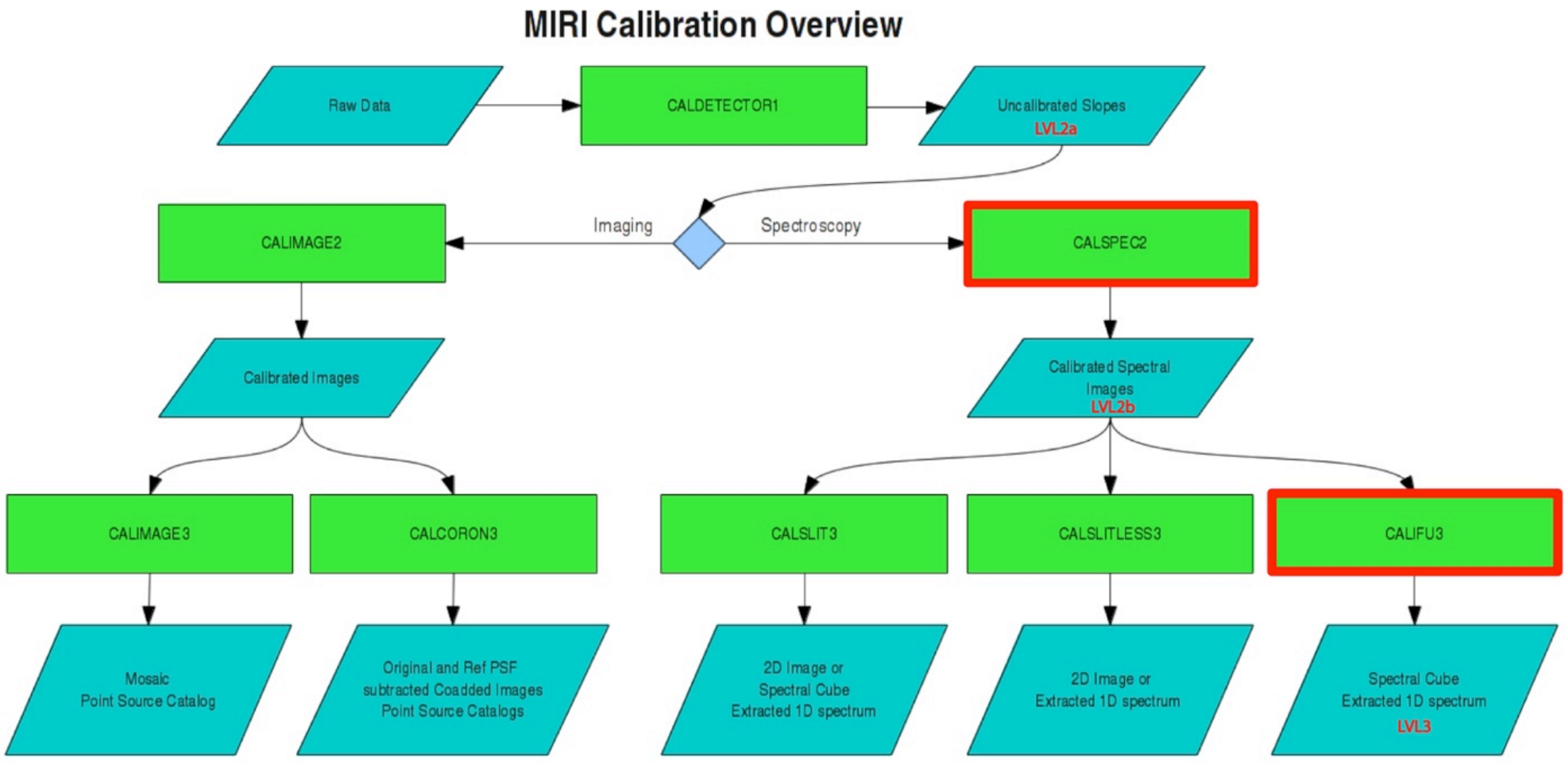

Figure 2 – Overview of the MIRI calibration process[5]. The reduction stages are shown as green rectangles and the input/output products are shown as blue trapezoids. The relevant steps for the MRS are highlighted in red.

Two versions of the MRS pipeline will be available: The default (*Baseline*) mode, and the advanced (*Optimal*) mode. The Baseline mode is designed to produce high quality data useful for most scientific cases foreseen. The Optimal mode is expected to be run just in a few exceptional cases where the user needs to assess very specific issues of the dataset.

Figure 3 and Figure 4 show the schematic description of the CALSPEC2 and CALSPEC3 steps for the Baseline mode. Depending on the spatial extension of the source (point-like or extended), different corrective steps are needed to account for diffraction effects, and the coarse spatial sampling in the slicing direction of the IFU. To provide a generic pipeline that runs independently for all sources, a decisive step will be implemented, in CALSPEC3, to determine autonomously the spatial shape of the source (while a user input allows for a customized run of the pipeline). Different branches of the pipeline will then be executed. All steps and calibration files have been implemented based on the results of the different MIRI test campaigns at Rutherford Appleton Laboratory (RAL) and Goddard Space Flight Center[9].

## 2.1 CALSPEC2

### 2.1.1 Assign WCS and distortion information

The first step of CALSPEC2 is to associate the exposure processed by CALDETECTOR1 with the reference data (distortion files) necessary to convert detector coordinates to output coordinates and wavelength (see e.g. [6] and [7]). Each sub-band of the MRS has a corresponding distortion file.

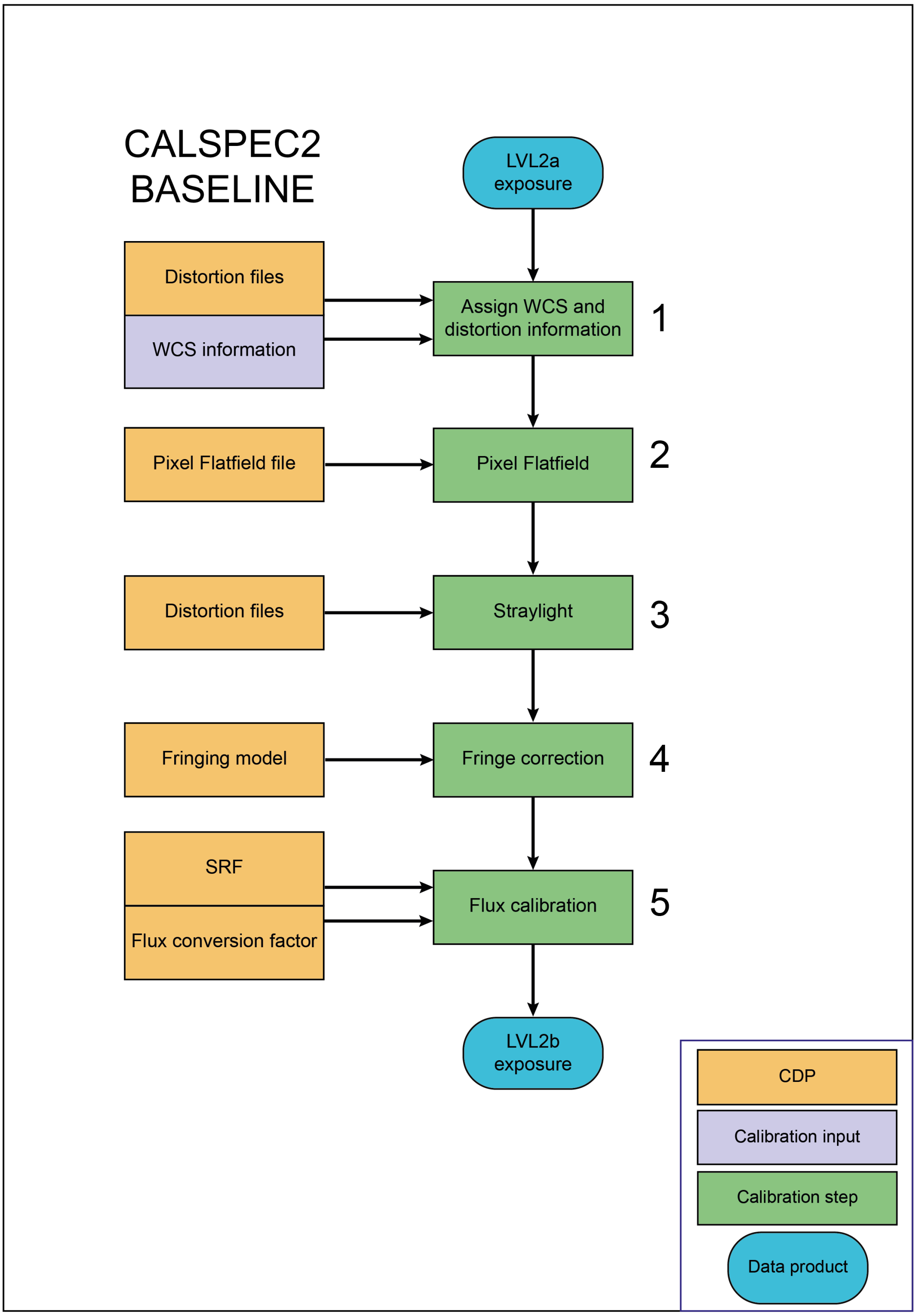

Figure 3 – Schematic description of CALSPEC2 Baseline

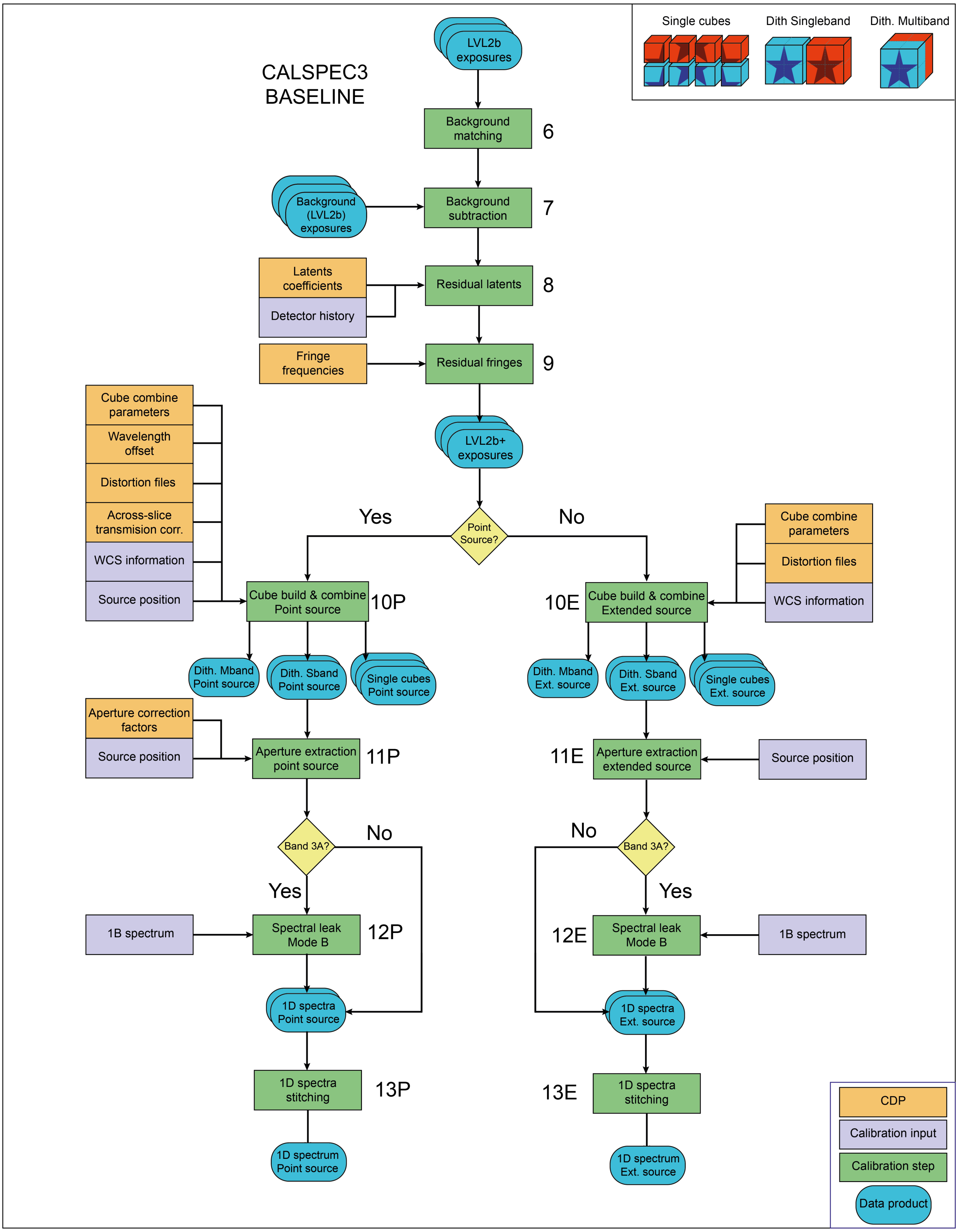

Figure 4 – Schematic description of CALSPEC3 Baseline

### 2.1.2 Pixel flatfield

This step corrects the pixel-to-pixel variations in gain at the detector, dividing by an array with the ratio of these individual gains over the average gain.

### 2.1.3 Stray-light

Testing of the Flight Module at RAL showed the presence of straylight in the MRS data, most likely due to scattered light from the spectrometer main optics (SMO). This stray-light adds to the sky and telescope background, and needs to be removed. This step of the pipeline removes the stray-light by modeling the distribution of light on the detector image, and subtracting it from the science image.

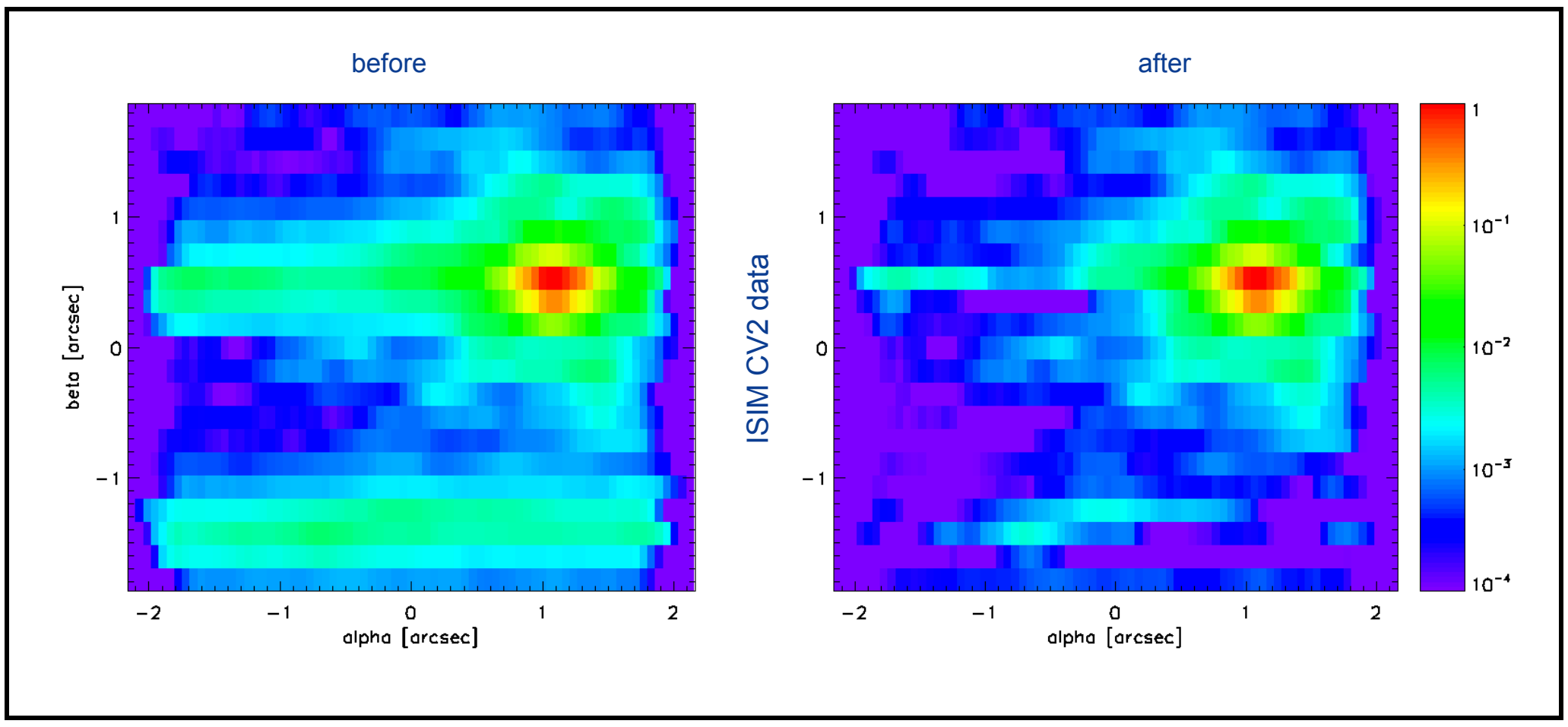

Figure 5 – Comparison of reconstructed images (collapsed cubes in wavelength direction) before and after stray-light removal for a point source, shown for sub-band 1A.

### 2.1.4 Fringe correction

The MRS data are affected by spectral fringes (periodic gain modulations caused by standing waves between parallel surfaces in the optical path) produced by the detector substrates, and the dichroic beam splitters, which act as Fabry-Pérot etalons. The MRS fringes have been modeled fitting a sinusoidal function to the normalized spectral response to a spatially unresolved source, and using test data to measure the fringe frequencies and maximum amplitudes. This step of the pipeline divides the detector-level data by this modeled fringe image.

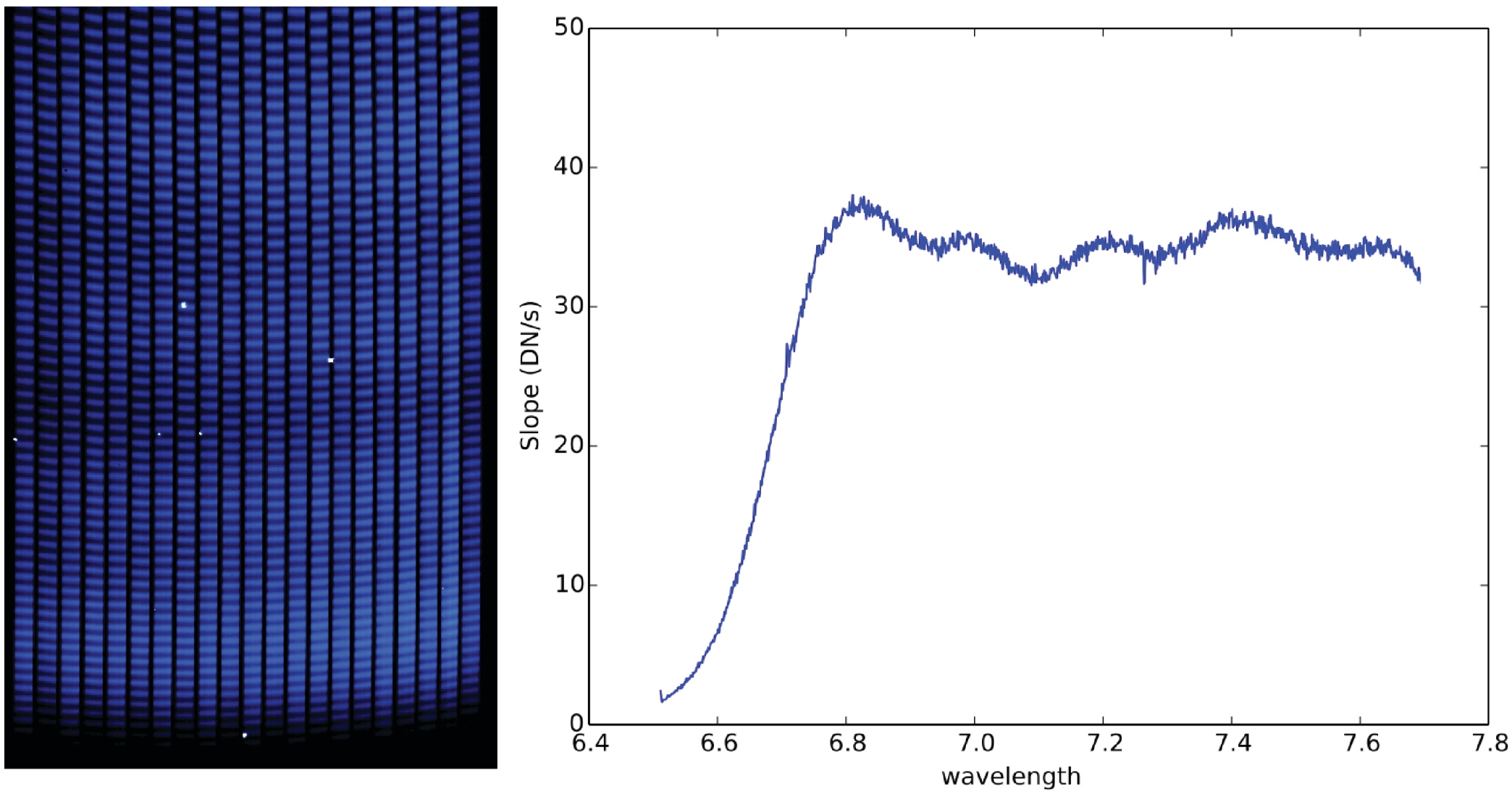

Figure 6 – *Left:* Zoom into an MRS exposure, highlighting the fringes on the detector image. *Right:* Spectrum of a slice from the same exposure.

### 2.1.5 Flux calibration

This step performs the spectro-photometrical calibration of the slope image. First, the variations in effective Photon Conversion Efficiency (PCE) throughout the detector image are compensated by means of the Spectral Response Function (SRF). The compensated variations in PCE include those caused by imperfections and inhomogeneities in the optics and the detector. After that division, the slope at each pixel is directly proportional to radiation energy per unit time, with a common constant of proportionality throughout the detector. The second step involves the application of proportionality constant, information which is present in the header, that allows converting the corrected pixel slopes to physical units of irradiance (e.g. W $m^{-2}$). The pipeline also includes a plane in the SRF with the angular area in the sky of the pixels. This allows computing specific intensities in mJy $arcsec^{-2}$.

## 2.2 CALSPEC3

### 2.2.1 Background matching

The first step of CALSPEC3 is matching the background of the different exposures in the association. If the user wants to run the pipeline for a single exposure, this step is not needed. Background matching equalizes the background level across dithered/mosaicked/spectrally overlapping exposures; and matches them to the background level of the exposure with the best estimation of background. It is achieved by adding a function of coordinates and wavelength (the Background Matching Function, BMF) to each slope image in an association. This addition yields a common background throughout the covered FOV and wavelength range, and across exposures. The algorithm reduces the number of exposures necessary to do background subtraction, in the case of multiple associated exposures, and minimizes background structure in mosaics. It also corrects for any background gradient present. If small, changing, gradients of the background are present over the FOV of MRS, it is possible to define a BMF that is a linear function of the slice coordinates. For a background with a smooth variation over wavelength (without spectral lines), the BMF would include coefficients with a slow dependency on wavelength.

### 2.2.2 Background subtraction

Background subtraction is accomplished subtracting the mean/median level of pixels slopes measured in a background (off-target/clear-sky) exposure. This is done for a number of spectral bins, to account for the wavelength-dependence of the background. The background level may be allowed to vary across the FOV, in which case the background has to be fitted in the off-target exposure (e.g. to some low degree polynomial), and then subtracted from the science image. Both the *background* and *science* exposures used in this step have gone through CALSPEC2, and are LEVEL 2b calibrated slope images.

### 2.2.3 Residual latents

The latency of the detector, i.e. its tendency to produce residual images caused by previous exposures, is corrected in the CALDETECTOR1 step of the pipeline, prior to CALSPEC2. Nonetheless, some latents (such as those generated by saturating signals) will be very hard to predict perfectly a priori, and may have not been completely corrected for in CALDETECTOR1. The pipeline will correct for those *residual* latents in this step. The algorithm will use a Calibration Data Product (CDP) with time-decay constants, and will need information on the previous state and history of the detector.

### 2.2.4 Residual fringes

The wavelength offset effect (see Section 2.10) may create residual fringes in some observations. The fringing patterns are wavelength and illumination-angle dependent, hence also depend on the surface brightness distribution of the source. For point source observations, we expect fringes at a level of a few percent. For extended source observations, all fringes should have been removed in CALSPEC2. However, based on the sensitivity of MIRI, it is likely that extended source observations include a weak point source in the background that the proposers were unaware of. This background source would produce residual fringes that need to be corrected here. This step corrects residual fringes at detector (2D) level, before cube building. The algorithm loops over the slices, then over spatial pixels per slice. At each position, the 1D spectrum will be extracted, corrected, and written back.

### 2.2.5 Cube build and combine

This step takes calibrated 2D slope images and produces 3D spectral cubes[7,8]. The IFU distorted and disjointed 2D spectra are corrected for distortion, and put back together into a rectangular cube with three orthogonal axes, two spatial and one spectral with regular sampling in the three axes (Figure 7).

These cubes are then combined to create dithered single-band cubes (the combination in spatial coordinates of two or more cubes, of a same sub-band), and dithered multi-band cubes (the combination in wavelength coordinates of two or more cubes, or dithered single-band cubes, of different sub-bands). The different outputs are illustrated in Figure 8. The FOV of the Dithered Single-band cube should be large enough not to crop the observed FOV of the combined exposures at any wavelength. The spectral range of the combined Dithered Multi-band cube will depend on the bands to be merged. All these "cubes" are created based on an association table of the relevant parameters for each combination of cubes.

The basics of cube creation and combination are the same for the point source branch and the extended source branch. The main difference between both branches is that, for the point source data, some additional corrections need to be applied while building the cubes. These corrections are "Wavelength offset" and "Across-slice transmission correction", and are dependent on the source position on the detector:

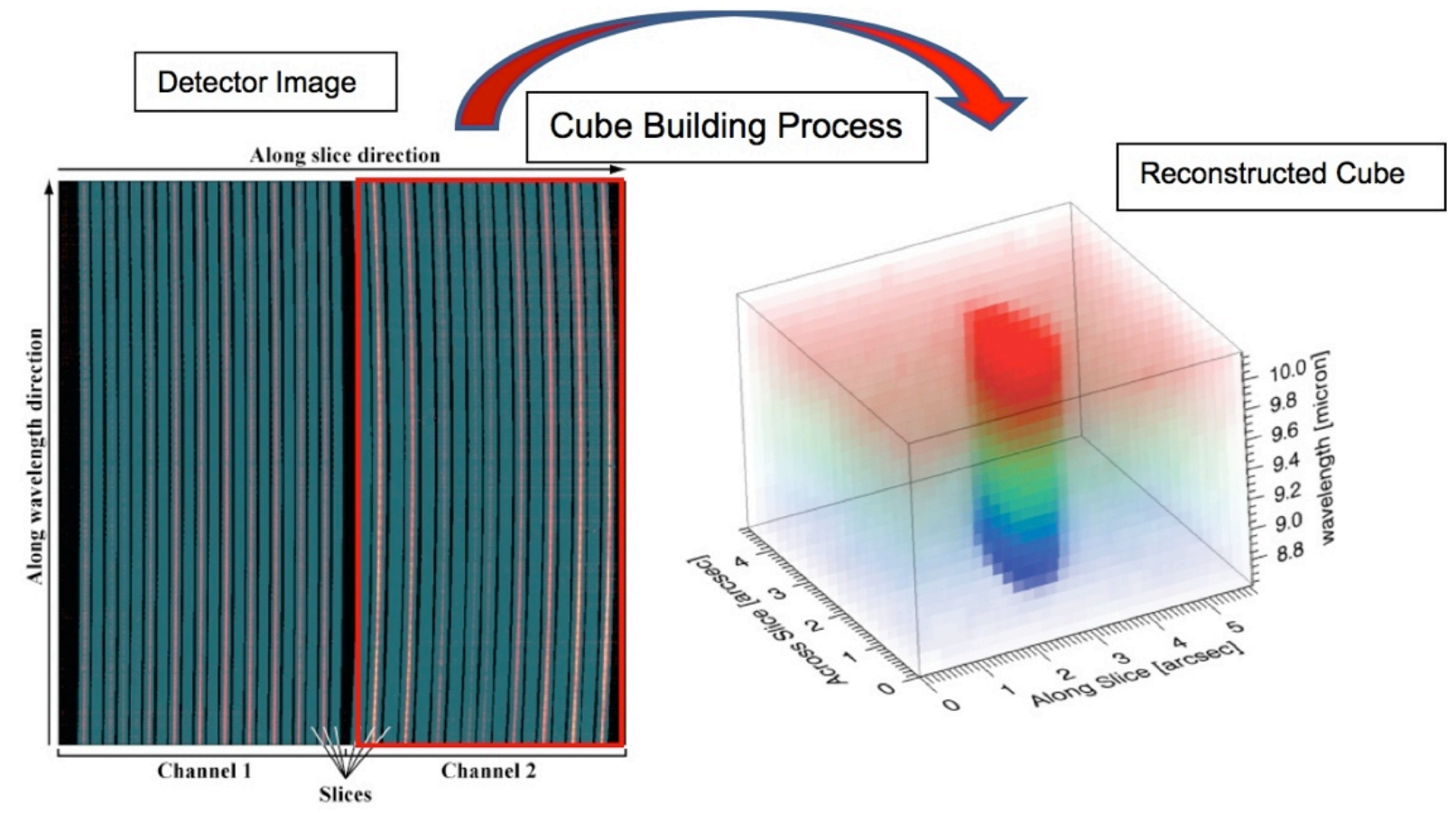

Figure 7 – Schematic of cube building. The spectra of each slice on the detector are undistorted and reorganized into a 3D matrix ("cube"). The distortion maps link the detector coordinates to sky coordinates and wavelength for each pixel.

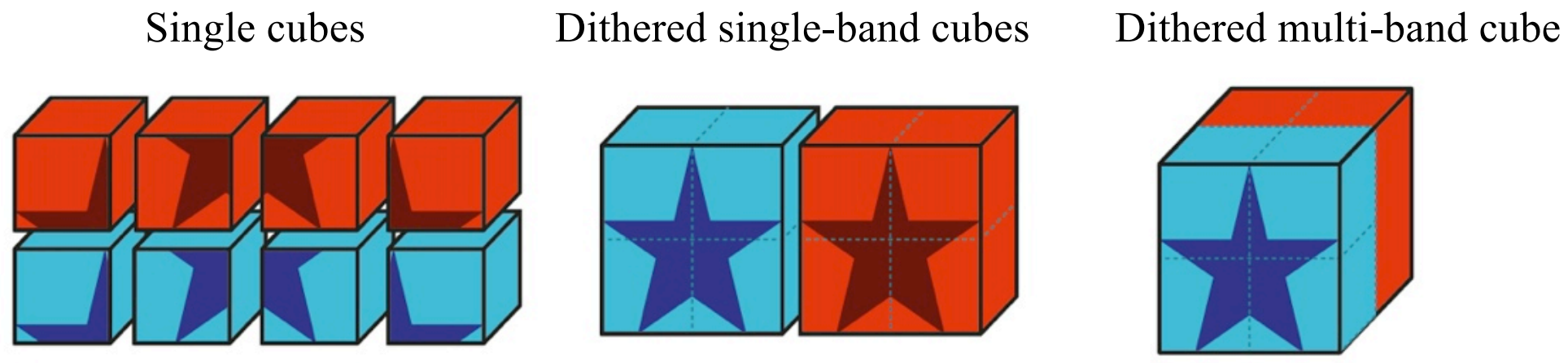

Figure 8 – Conceptual description of the different outputs of cube building for each branch (point source and extended source). In this example, cube build would be working on four pointings with two sub-bands (red and blue) per pointing.

*Wavelength offset*: The MRS spectrum of a point source is slightly shifted in wavelength depending on the position of the source relative to the slice, in the across-slice dimension. In the case of an extended source, which can always be understood as a continuum distribution of point-like sources over the FOV, this effect smooths out.

*Across slice transmission correction*: The slicer mirror breaks the wavefronts coming from the Spectrometer Pre-Optics (SPO), generating diffraction losses. The diffraction losses at each slice depend on the across-slice distance of the source to the slice center. For point-like sources, the losses of the slices do not add up to a constant, and a scaling-factor correction to the spectrum of each slice must be applied. For an extended source, which can be regarded as a continuum distribution of point sources over the FOV, the losses of all slices add up to a constant value that is factored in the spectrophotometric calibration.

### 2.2.6 Aperture extraction.

Aperture extraction is done for each Dithered Single-band cube, providing a 1D spectrum per band. The algorithm performs a collapse of the 3D spectrum in a spatial aperture containing the source, analogous to the *classical* aperture extraction in photometry, but applied at each $\lambda$=constant plane of the cube. In the point source branch, the aperture correction step takes into account the MRS PSF.

#### 2.2.7 Spectral leak.

The MRS suffers from a minor spectral leak, by which a small fraction (~2-3%) of the light at 6.1 µm (sub-band 1B) is leaked into sub-band 3A (at ~12.2 µm), because of inefficient order selection filtering. Hence, this step is applied only to spectra in band 3A. The algorithm works similarly for the point source and extended source branches. This step models the leakage using the spectrum at 6.1 µm from band 1B, and subtracts that model from the 3A spectrum. The correction is applied on the extracted 1D spectra

#### 2.2.8 1D spectra stitching.

The last step of CALSPEC3 is stitching all the 1D spectra to produce one single spectrum including all sub-bands observed. The algorithm works similarly for both point source and extended source branches.

## 3. SUMMARY

The extensive MIRI testing campaigns performed in Europe and the USA allowed us to accurately model and characterize all the intricacies of the MRS and its data. The complexity of the MRS required a very specialized pipeline, with some specific steps not present in other pipelines of JWST instruments. The MRS Calibration Pipeline, overviewed in this paper, has been designed to guarantee the highest scientific return from all MRS observations during, and after, JWST operations.